\documentclass[preprint,aps,showpacs,nofootinbib,preprintnumbers,amsmath,amssymb]{revtex4-1}
\usepackage{epsfig}
\usepackage{subfigure}
\usepackage{multirow}
\usepackage{dcolumn}
\usepackage{bm}
\usepackage[usenames ,dvipsnames]{xcolor}
\usepackage{slashed}
\usepackage{float}
\usepackage{subfigure}
\usepackage{graphicx,color}
\usepackage{amsmath}
\usepackage{slashed}
\usepackage{nonfloat}
\begin{document}
\title{  Semileptonic decays of $\Lambda_c^+$ in dynamical approaches}

\author{C.Q. Geng$^{1,2,3,4}$, Chong-Chung Lih$^{5}$, Chia-Wei Liu$^{3}$ and Tien-Hsueh Tsai$^{3}$}
\affiliation{
$^{1}$School of Fundamental Physics and Mathematical Sciences, Hangzhou Institute for Advanced Study, UCAS, Hangzhou 310024, China \\
$^{2}$International Centre for Theoretical Physics Asia-Pacific, Beijing/Hangzhou, China \\
$^{3}$Department of Physics, National Tsing Hua University, Hsinchu 300, Taiwan\\
$^{4}$Physics Division, National Center for Theoretical Sciences, Hsinchu 300, Taiwan\\
$^{5}$Department of Optometry,  Central Taiwan University of Science and Technology, Taichung 406, Taiwan 
}\date{\today}

\begin{abstract}
We study the semileptonic decays of $\Lambda_c^+ \to \Lambda(n)\ell^+ \nu_{\ell}$ in two relativistic dynamical approaches of the light-front constituent quark model (LFCQM) and  MIT bag model (MBM).  By considering the Fermi statistic between quarks and  determining spin-flavor structures  in baryons
 along with the helicity formalism in the two different dynamical models, we  calculate the branching ratios (${\cal B}$s) and averaged asymmetry parameters ($\alpha$s) in the decays. Explicitly, we find that  ${\cal B}( \Lambda_c^+ \to \Lambda e^+ \nu_{e})=(3.36\pm0.87,3.48)\%$ and
${\alpha}( \Lambda_c^+ \to \Lambda e^+ \nu_{e})=(-0.97\pm0.03,-0.83)$ in (LFCQM, MBM), 
in comparison  with the data of ${\cal B}( \Lambda_c^+ \to \Lambda e^+ \nu_{e})=(3.6\pm0.4)\%$ and
${\alpha}( \Lambda_c^+ \to \Lambda e^+ \nu_{e})=-0.86\pm 0.04$ given by the Particle Data Group, respectively.
We also predict that ${\cal B}(  \Lambda_c^+ \to n e^+ \nu_{e})=(0.57\pm0.15, 3.6\pm1.5)\times 10^{-3}$ and
${\alpha}( \Lambda_c^+ \to n e^+ \nu_{e})=(-0.98\pm0.02,-0.96\pm0.04)$  in LFCQM with two different scenarios for the momentum distributions of quarks 
in the neutron, and ${\cal B}(  \Lambda_c^+ \to n e^+ \nu_{e})= 0.279\times 10^{-2}$ and ${\alpha}( \Lambda_c^+ \to n e^+ \nu_{e})=-0.87$ in MBM, which could be tested by the ongoing experiments at BESIII, LHCb and  BELLEII. 
\end{abstract}
\maketitle

\section{introduction}
Recently, the LHCb Collaboration has published the newest precision measurements on the anti-triplet charmed baryon lifetimes~\cite{Aaij:2019lwg}, given by
\begin{eqnarray}
\label{q0}
\tau_{\Lambda_c^+}&=&203.5\pm 1.0\pm 1.3\pm 1.4 \text{ fs}\,, \nonumber\\
\tau_{\Xi_c^+}&=&456.8\pm 3.5 \pm 2.9 \pm 3.1 \text{ fs}\,, \nonumber \\ 
\tau_{\Xi_c^0}&=&154.5 \pm 1.7 \pm 1.6 \pm 1.0 \text{ fs}\,. 
\end{eqnarray}
Surprisingly, the lifetime of $\Xi_c^0$  given by LHCb magnificently deviates from the previous value of $\tau_{\Xi_c^0}=112^{+13}_{-10}\text{ fs}$ in PDG~\cite{Tanabashi:2018oca}. Meanwhile, the Belle Collaboration has measured the absolute branching ratios of 
${\cal B} (\Xi_c^0 \to \Xi^- \pi^+)=(1.8\pm0.5 )\%$~\cite{Li:2018qak} and
${\cal B} (\Xi_c^+ \to \Xi^- \pi^+\pi^+)=(2.86 \pm 1.21 \pm 0.38)\%$~\cite{Li:2019atu}, which are the golden modes to determine other $\Xi_c^{0,+}$ decay channels, respectively.
It is clear that we are now witnessing a new era of charm physics. One can expect that there will be more and more new experimental data and precision measurements in the near future, which  are also the guiding light for people to explore new physics beyond the standard model.   

There have been recently many works  discussing the anti-triplet charm baryon decays.
Because of the complicated baryon structures, particularly the large non-perturbative effects in the quantum chromodynamics (QCD), 
it is very hard to calculate the baryonic decay amplitudes from first principles. 
In the literature, people use the flavor symmetry of $SU(3)_f$ 
to analyze various charmed baryon decay processes, such as  semi-leptonic, two-body and three-body non-leptonic decays, to obtain reliable results~\cite{Geng:2017mxn,Savage:1989qr,Savage:1991wu,zero,Geng:2018bow,He:2018joe,He:2018php,Wang:2017azm,Geng:2018plk,Geng:2018rse,Geng:2018upx,Geng:2019awr,Geng:2019bfz,Geng:2019xbo,Grossman:2019xcj,Lu:2016ogy,Wang:2017gxe,Cen:2019ims,Hsiao:2019yur,Roy:2019cky,Jia:2019zxi}.
However, the $SU(3)_f$ symmetry is an approximate symmetry, resulting in about $10\%$ of errors for the predictions   naturally. 
In order to have more precise calculations,
 we need a dynamical QCD model to understand each process.
%
For simplicity, we only discuss the semi-leptonic processes, which involve purely the factorizable effects without the non-factorizable ones.
 In particular, we focus on the  $\Lambda_c^+$ semi-leptonic decays in this work.
There are several theoretical analyses and lattice QCD (LCCQ) calculations on the charmed baryon semi-leptonic decays with different dynamical models 
in the literature~\cite{Cheng:1995fe,Zhao:2018zcb,PerezMarcial:1989yh,Faustov:2016yza,Gutsche:2014zna,Gutsche:2015rrt,Meinel:2016dqj,Meinel:2017ggx}. 
In this paper, we will mainly use the light-front (LF) formalism to study the decays and check the results  
in the MIT bag model (MBM) as comparisons.

The LF formalism is considered as a consistent relativistic approach, which has been very  successful  
in  the mesonic  sectors~\cite{Schlumpf:1992ce,Zhang:1994ti}. 
Due to this success, it has been extended  to other systems, such as those involving the heavy mesons, pentaquarks and so on~\cite{Cheng:2003sm,Cheng:2004cc,Chang:2019obq,Zhao:2018mrg,Xing:2018lre,Cheng:2017pcq,Geng:1997ws,Lih:1999it,Geng:2000fs,Geng:2001de,Geng:2003su,Geng:2016pyr}.
In addition, the bottom baryon to charmed baryon nonleptonic decays in the LF approach have been done in Refs.~\cite{Chua:2018lfa,Chua:2019yqh}.
For a review on the non-perturbative nature in the equation of motion
and  QCD vacuum structure for the LF constituent quark model (LFCQM), one can refer to the article of Ref.~\cite{Zhang:1994ti}.
The  advantage of LFCQM  is that
the commutativity of the LF Hamiltonian and  boost generators  provide us with a good convenience to calculate the wave-function in different inertial frames because of the recoil effect.

MBM is a QCD inspired phenomenological model. In MBM,  a baryon is described as three free quarks with current masses confined in a spherical bag with a bag size R. This simple and intuitive picture helps people to deal with the interactions between hadrons as well as their mass spectra. 
The authors in Ref.~\cite{PerezMarcial:1989yh} have calculated all $c\to s/d$ baryonic transition form factors at zero-recoiled points and discussed both the monopole and dipole behaviors, and others in Refs.~\cite{Cheng:1991sn,Cheng:2018hwl,Zou:2019kzq} have further combined MBM with the pole model and current algebra to predict various charmed baryon non-leptonic decays. 

This paper is organized as follows. We present our formal calculations of the baryonic transition form factors for LFCQM and MBM in Secs.~II and
III, respectively.
  We show our numerical results of the form factors, branching ratios and averaged asymmetry parameters in Sec. IV.
  We also  compare our results  with those in the literature.  In Sec.~V, we give our   conclusions.

\section{Baryonic transition form factors in LFCQM}
\subsection{Vertex function of baryon}
In LFCQM, a baryon with its momentum $P$ and spin $S$ as well as the z-direction projection of spin $S_z$ are considered as a bound state of three constitute quarks.
As a result, the baryon state can be expressed by~\cite{Zhang:1994ti,Schlumpf:1992ce,Cheng:2004cc,Ke:2007tg,Ke:2012wa,Ke:2019smy}
\begin{eqnarray}
\label{eq1}
	&|{\bf B}&,P,S,S_z\rangle=\int\{{d^3{\tilde{p}_1}}\}\{{d^3{\tilde{p}_2}}\}\{{d^3{\tilde{p}_3}}\}2(2\pi)^3\frac{1}{\sqrt{P^+}}\delta^3(\tilde{P}-\tilde{p}_1-\tilde{p}_2-\tilde{p}_3) \nonumber \\
	&\times& \sum_{\lambda_1,\lambda_2,\lambda_3}\Psi^{SS_z}(\tilde{p}_1,\tilde{p}_2,\tilde{p}_3,\lambda_1,\lambda_2,\lambda_3)C^{\alpha\beta\gamma}F_{abc}|q_{\alpha}^{a}(\tilde{p}_1,\lambda_1) q_{\beta}^{b}(\tilde{p}_2,\lambda_2)q_{\alpha}^{a}(\tilde{p}_3,\lambda_3) \rangle,
	\label{baryon}
\end{eqnarray}
where  $\Psi^{SS_z}(\tilde{p}_1,\tilde{p}_2,\tilde{p}_3,\lambda_1,\lambda_2,\lambda_3)$  is
 the vertex function, which can be formally solved from the Bethe-Salpeter equations by the Faddeev decomposition method,
$C^{\alpha\beta\gamma}$ and $F_{abc}$ are the color and flavor factors, $\lambda_i$  and $\tilde{p}_i$ with $i=1,2,3$  are the LF helicities and
3-momentua of the on-mass-shell constituent quarks, defined as
\begin{eqnarray}
\tilde{p}_{i}=(p_i^+,p_{i\perp})\,, ~p_{i\perp}=(p_i^1,p_i^2) \,,~ p_i^-=\frac{m_i^2+p_{i\perp}^2}{p_i^+} \,,
\end{eqnarray}
and
\begin{eqnarray}
	&&{d^3\tilde{p}_i}\equiv \frac{dp_i^+d^2p_{i\perp}}{2(2\pi)^3}\,, ~ \delta^3(\tilde{p})=\delta(p^+)\delta^2(p_{\perp})\,,
	 \nonumber\\
	&&|q_\alpha^a(\tilde{p},\lambda)\rangle=d^{\dagger a}_{\alpha}(\tilde{p},\lambda)|0\rangle\,,
	 ~\{d_{\alpha'}^{a'}(\tilde{p'},\lambda'),d^{\dagger a}_{\alpha}(\tilde{p},\lambda)\}=2(2\pi)^{3}\delta^3(\tilde{p'}-\tilde{p})\delta_{\lambda'\lambda}\delta_{\alpha'\alpha}\delta^{a'a}\,,
	\label{state}
\end{eqnarray}
respectively.
To describe the internal motions of the constituent quarks, we introduce the  kinematic variables of $(q_{\perp},\xi)$,
 $(Q_{\perp},\eta)$ and $P_{tot}$, given by
\begin{eqnarray}
P_{tot}&=&\tilde{P}_1+\tilde{P}_2+\tilde{P}_3, \qquad \xi=\frac{p_1^+}{p_1^++p_2^+}, \qquad \eta=\frac{p_1^++p_2^+}{P_{tot}^+}\,,
 \nonumber \\
q_{\perp}&=&(1-\xi)p_{1\perp}-\xi p_{2\perp},\quad  Q_{\perp}=(1-\eta)(p_{1\perp}+p_{2\perp})-\eta p_{3\perp} \,,
\label{Lkin}
\end{eqnarray}
where $(q_{\perp},\xi)$ characterize the relative motion between the first and second quarks, while $(Q_{\perp},\eta)$  the third and other two quarks. 
The invariant masses of  $(q_{\perp},\xi)$ and $(Q_{\perp},\eta)$ systems are represented by~\cite{Schlumpf:1992ce}
\begin{eqnarray}
M_3^2=\frac{q_\perp^2}{\xi(1-\xi)}+\frac{m_1^2}{\xi}+\frac{m_2^2}{1-\xi}\,, \nonumber\\
M^2=\frac{Q_\perp^2}{\eta(1-\eta)}+\frac{M_3^2}{\eta}+\frac{m_3^2}{1-\eta}\,,
\end{eqnarray}
respectively.
Unlike Refs.~\cite{Ke:2007tg,Ke:2012wa,Ke:2019smy}, which treat the diquark as a point like object or spectator, 
we consider the three constituent quarks in the baryon independently with suitable quantum numbers satisfying the Fermi statistics
to have a correct baryon bound state system. 
The vertex function of $\Psi^{SS_z}(\tilde{p}_1,\tilde{p}_2,\tilde{p}_3,\lambda_1,\lambda_2,\lambda_3)$ in Eq.~(\ref{eq1}) 
can be written as~\cite{Lorce:2011dv,Zhang:1994ti,Schlumpf:1992ce}
\begin{eqnarray}
\Psi^{SS_z}(\tilde{p}_1,\tilde{p}_2,\tilde{p}_3,\lambda_1,\lambda_2,\lambda_3)&=&\Phi(q_\perp,\xi,Q_\perp,\eta)\Xi^{SS_z}(\lambda_1,\lambda_2,\lambda_3)\,,
\end{eqnarray}
where $\Phi(q_\perp,\xi,Q_\perp,\eta)$ is
 the momentum distribution of constituent quarks  and $\Xi^{SS_z}(\lambda_1,\lambda_2,\lambda_3)$ represents
 the momentum-dependent  spin wave function, given by 
 \begin{eqnarray}
\Xi^{SS_z}(\lambda_1,\lambda_2,\lambda_3)&=&\sum_{s_1,s_2,s_3}\langle\lambda_1|R^{\dagger}_1|s_1\rangle\langle\lambda_2|R^{\dagger}_2|s_2\rangle\langle\lambda_3|R^{\dagger}_3|s_3\rangle \bigg\langle\frac{1}{2}s_1,\frac{1}{2}s_2,\frac{1}{2}s_3\bigg|SS_z\bigg\rangle\,,
\end{eqnarray}
with
 $\big\langle\frac{1}{2}s_1,\frac{1}{2}s_2,\frac{1}{2}s_3\big|SS_z\big\rangle$  the usual $SU(2)$ Clebsch-Gordan coefficient, and  $R_i$ the well-known Melosh transformation,  which corresponds to the $i$th constituent quark, expressed by
\begin{eqnarray}
&&R_1=R_M(\eta,Q_\perp,M_3,M)R_M(\xi,q_\perp,m_1,M_3)\,, \nonumber\\
&&R_2=R_M(\eta,Q_\perp,M_3,M)R_M(1-\xi,-q_\perp,m_2,M_3) \,,\nonumber\\
&&R_3=R_M(1-\eta,-Q_\perp,m_3,M)\,,
\end{eqnarray}
with 
\begin{eqnarray}
R_M(x,p_{\perp},m,M)&=&\frac{m+xM-i\vec{\sigma}\cdot(\vec{n}\times \vec{q})}{\sqrt{(m+xM)^2+q_\perp^2}},
\end{eqnarray}
where $\vec{\sigma}$ stands for the Pauli matrix and $\vec{n}=(0,0,1)$. 
This is the generalization of the Melosh transformation from two-particle systems,
which can be derived from the transformation property of angular momentum operators~\cite{Polyzou:2012ut,Schlumpf:1992ce}.
We further represent the LF kinematic variables $(\xi,q_\perp)$ and $(\eta,Q_\perp)$ in the forms of the ordinary 3-momenta ${\bf q}$ and  ${\bf Q}$: 
\begin{eqnarray}
	&&E_{1(2)}=\sqrt{{\bf q}^2+m_{1(2)}^2} \,,\quad E_{12}=\sqrt{{\bf Q}^2+M_3^2}\,, \quad E_3=\sqrt{{\bf Q}^2+m_3^2}\,, \nonumber \\
	&&q_z=\frac{\xi M_3}{2}-\frac{m_1^2+q_\perp^2}{2M_3\xi}\,, \quad Q_z=\frac{\eta M}{2}-\frac{M_3^2+Q_\perp^2}{2M\eta} \,,
	\label{3kin}
\end{eqnarray}
to get more clear physical pictures of the momentum distribution wave functions.

It is known that the exact momentum wave function cannot be solved from the  first principle currently due to the lack of knowledge 
about the  effective potential in the three-body system in QCD.
Hence, we choose the phenomenological Gaussian type wave function with suitable shape parameters to include the diquark clustering effects 
in $\Lambda_c^+$ and $\Lambda$ baryons~\cite{Ke:2019smy,Schlumpf:1992ce}.
The baryon spin-flavor-momentum wave function $F_{abc}\Psi^{SS_z}(\tilde{p}_1,\tilde{p}_2,\tilde{p}_3,\lambda_1,\lambda_2,\lambda_3)$ should be totally symmetric under any permutations of quarks to keep the Fermi statistics. The spin-flavor-momentum wave functions of $\Lambda_c^+$, $\Lambda$ and $n$ are given by
\begin{eqnarray}
	&&|\Lambda_c\rangle=\frac{1}{\sqrt{6}}[\phi_3\chi^{\rho3}(|duc\rangle-|udc\rangle)+\phi_2\chi^{\rho2}(|dcu\rangle-|ucd\rangle)+\phi_1\chi^{\rho1}(|cdu\rangle-|cud\rangle)]\,,
	\nonumber\\
		&&|\Lambda\rangle=\frac{1}{\sqrt{6}}[\phi_3\chi^{\rho3}(|duc\rangle-|uds\rangle)+\phi_2\chi^{\rho2}(|dsu\rangle-|usd\rangle)+\phi_1\chi^{\rho1}(|sdu\rangle-|sud\rangle)]\,,
		 \nonumber\\
		&&|n\rangle=\frac{1}{\sqrt{3}}\phi[\chi^{\lambda_3}|ddu\rangle+\chi^{\lambda_2}|dud\rangle+\chi^{\lambda_1}|udd\rangle]\,,
\end{eqnarray}		
respectively, where
\begin{eqnarray}		
		\chi^{\rho 3}_{\uparrow}&=&\frac{1}{\sqrt{2}}(|\uparrow\downarrow\uparrow\rangle-|\downarrow\uparrow\uparrow\rangle)\,,
		 \quad \chi^{\lambda 3}_{\uparrow}=\frac{1}{\sqrt{6}}(|\uparrow\downarrow\uparrow\rangle+|\downarrow\uparrow\uparrow\rangle-2|\uparrow\uparrow\downarrow\rangle)\,,
		 \nonumber\\
	\phi_3&=&{\cal N}\sqrt{\frac{\partial q_z}{\partial \xi}\frac{\partial Q_z}{\partial \eta}}e^{-\frac{{\bf Q}^2}{2\beta_{Q}^2}-\frac{{\bf q}^2}{2\beta_q^2}}\,, 	 
\end{eqnarray}
 and $\phi_{1(2)}$ has the form  by replacing $({\bf q},{\bf Q})$ with $({\bf q}_{1(2)},{\bf Q}_{1(2)})$  in $\phi_3$, 
 with ${\cal N}=2(2\pi)^3(\beta_q\beta_{Q}\pi)^{-3/2}$ and  $\beta_{q,Q}$ being the normalized constant
and   shape parameters, respectively.
Explicitly,  ${\bf q}_{1(2)}$ and ${\bf Q}_{1(2)}$  are given by
\begin{eqnarray}
\xi_{1(2)}&=&\frac{p_{2(3)}^+}{p_{2(3)}^++p_{(1)}^+}, \qquad \eta_{1(2)}=1-\frac{p_{1(2)}^+}{P_{tot}^+}\,, \nonumber \\
q_{1(2)\perp }&=&(1-\xi_{1(2)})p_{2(3)\perp}-\xi_{1(2)} p_{3(1)\perp},\nonumber \\
  Q_{1(2)\perp}&=&(1-\eta_{1(2)})(p_{2(3)\perp}+p_{3(1)\perp})-\eta_{1(2)} p_{1(2)\perp} \,.
\end{eqnarray}
 Here, the baryon state is normalized as
\begin{eqnarray}
\langle {\bf B}&,P',S',S'_z|{\bf B}&,P,S,S_z\rangle=2(2\pi)^3P^+\delta^3(\tilde{P'}-\tilde{P})\delta_{S_z'S_z}\,,
\label{baryonN}
\end{eqnarray}
resulting in
the normalization of the momentum wave function, given by
\begin{eqnarray}
\frac{1}{2^2(2\pi)^6}	\int d\xi_{(1,2)} d\eta_{(1,2)} d^2q_{(1,2)\perp}d^2Q_{(1,2)\perp} |\phi_{3(1,2)}|^2=1\,.
\end{eqnarray} 
We emphasize that the momentum wave functions of $\phi_i$ with the different shape parameters of $\beta_q$ and $\beta_{Q}$  describe the scalar diquark effects
 in $\Lambda_{(c)}$.  
 For the neutron, the momentum distribution functions are the same, $i.e.$
 $\phi=\phi_3(\beta_q=\beta_{Q})$, for any spin-flavor state due to the  isospin symmetry. 
 Note that there is no $SU(6)$ spin-flavor symmetry in $\Lambda_{(c)}$  
 even though the forms of these states are similar to those with  $SU(6)$. 
 
\subsection{Transition form factors}
The baryonic transition form factors  of the $V-A$  weak current are defined by
\begin{eqnarray}
&&\langle {\bf B}_f,P',S',S'_z|\bar{q}\gamma^{\mu}(1-\gamma_5)c|{\bf B}_i,P,S,S_z\rangle \nonumber\\
&&=\bar {u}_{{\bf B}_f}(P',S'_z)\bigg[\gamma^{\mu}f_1(k^2)-i\sigma^{\mu\nu}\frac{k_{\nu}}{M_{{\bf B}_i}}f_2(k^2)+\frac{k^{\mu}}{M_{{\bf B}_i}}f_3(k^2)\bigg]u_{{\bf B}_i}(P,S_z)\nonumber\\
&&-\bar {u}_{{\bf B}_f}(P',S'_z)\bigg[\gamma^{\mu}g_1(k^2)-i\sigma^{\mu\nu}\frac{k_{\nu}}{M_{{\bf B}_i}}g_2(k^2)+\frac{k^{\mu}}{M_{{\bf B}_i}}g_3(k^2)\bigg]\gamma_5u_{{\bf B}_i}(P,S_z)
\end{eqnarray}
where $\sigma^{\mu\nu}=\frac{i}{2}[\gamma^{\mu},\gamma^{\nu}]$ and $P'-P=k$.
We choose the frame such that $P^+$ is conserved ($k^+=0,k^2=-k_\perp^2$) to calculate the form factors to avoid  
other $x^{+}$-ordered diagrams in the LF formalism~\cite{Schlumpf:1992ce}.
The matrix elements of the vector and axial-vector currents at quark level correspond to three different lowest-order Feynman diagrams 
as shown in Fig.~1. 
\begin{figure}
 \begin{minipage}[h]{0.3\linewidth}
 	 	(a)
 	\centering
 	\includegraphics[width=2in]{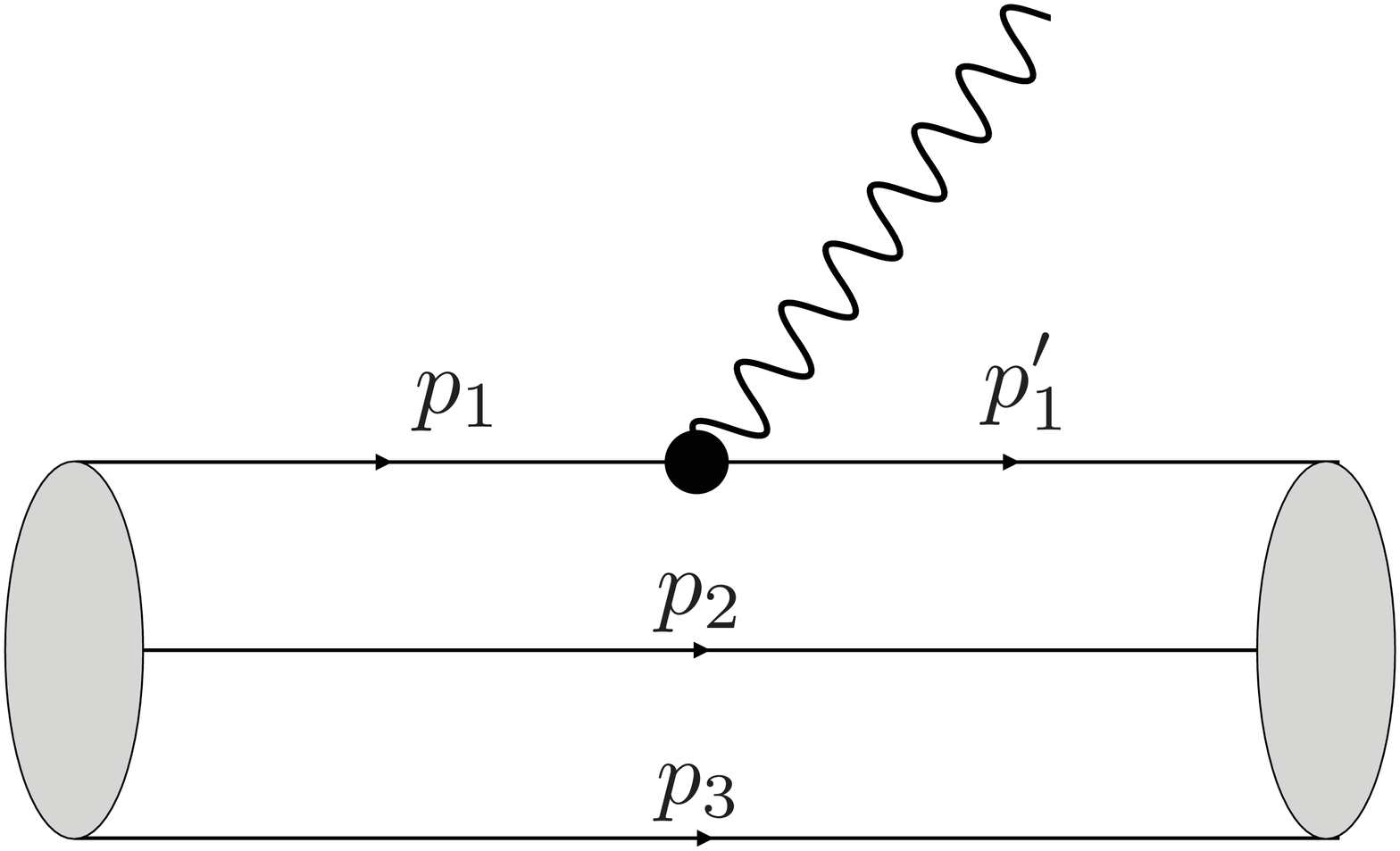}
 \end{minipage}
\begin{minipage}[h]{0.3\linewidth}
	(b)
 	\centering
\includegraphics[width=2in]{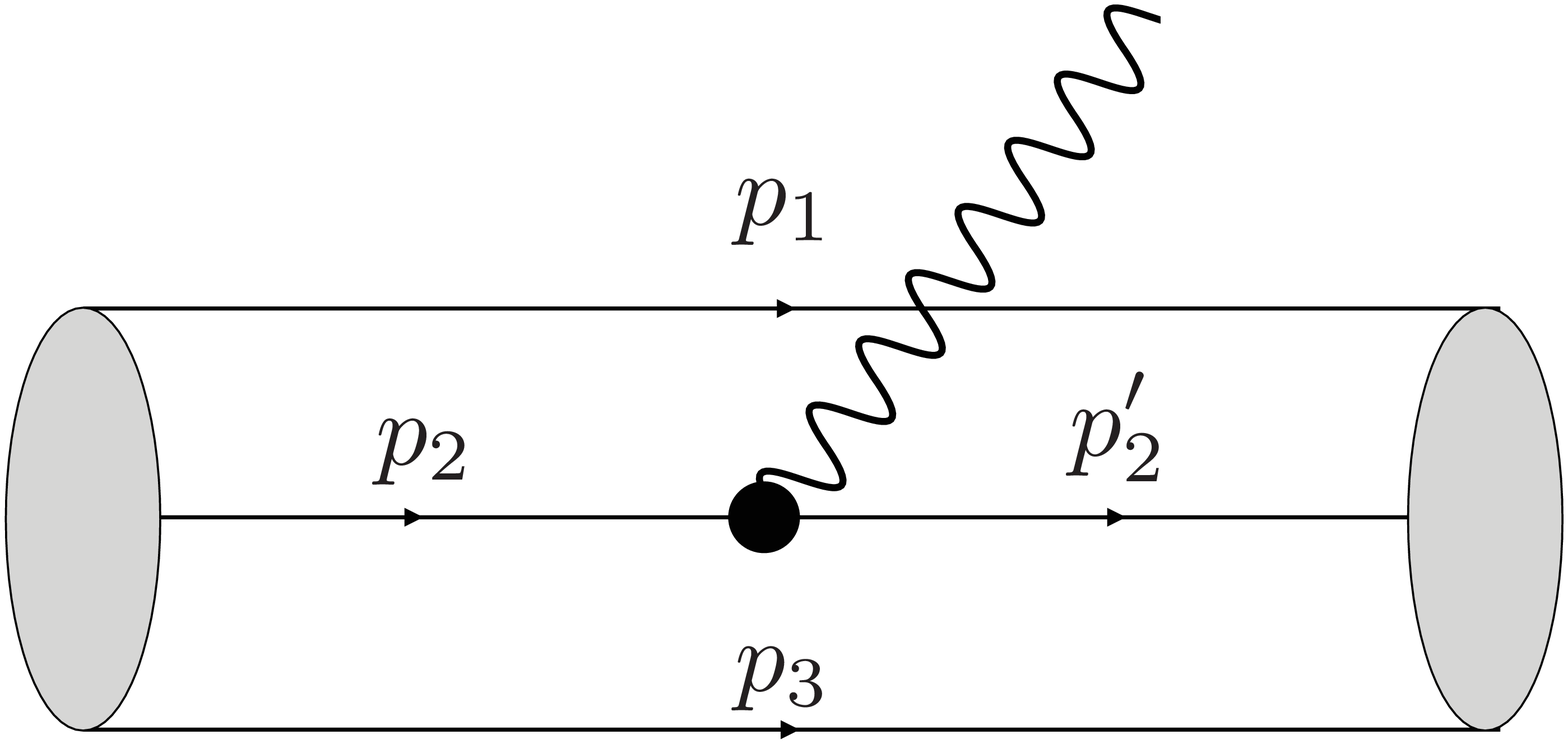}.
\end{minipage}
\begin{minipage}[h]{0.3\linewidth}
	(c)
 	\centering
\includegraphics[width=2in]{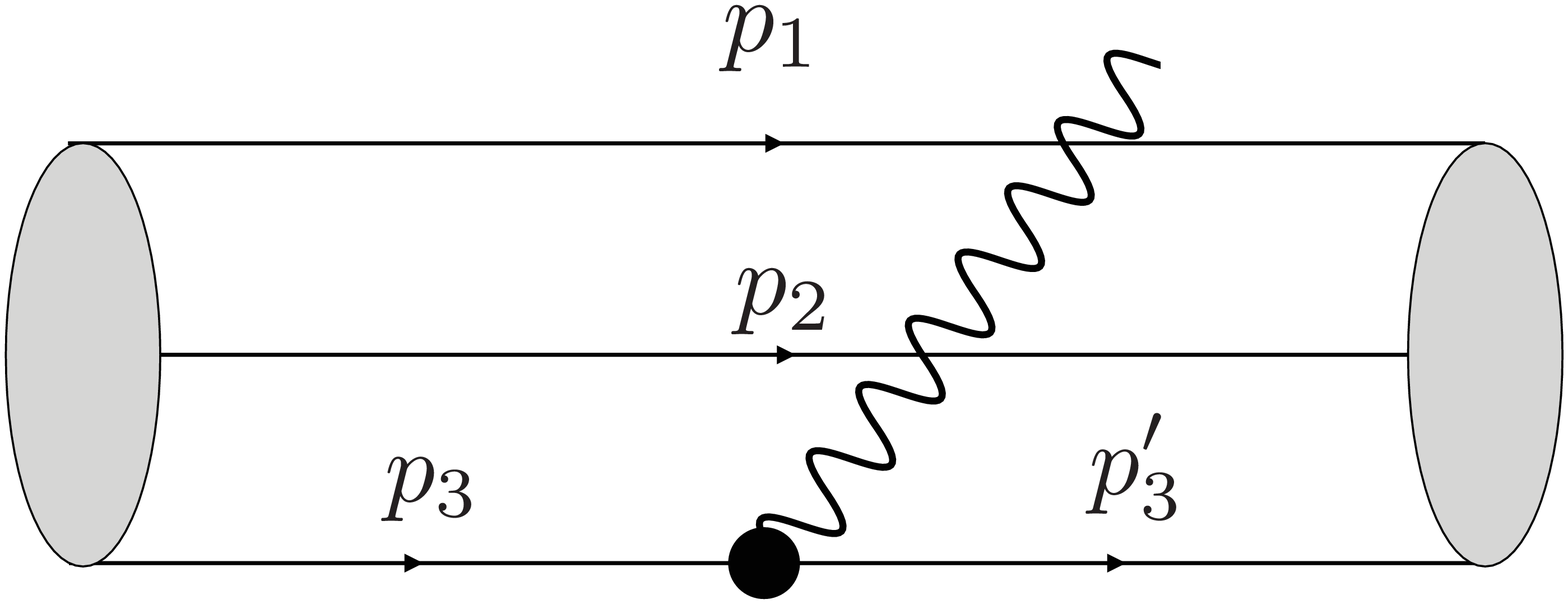}
\end{minipage}
\caption{Feynman diagrams for the baryonic weak transitions at the lowest order, where the sign of ``$\bullet$'' denotes the V-A  current vertex,
with  (a) $p'_1-p_1=k$, (b) $p'_2-p_2=k$ and (c) $p'_3-p_3=k$.}
\end{figure}
Since the spin-flavor-momentum wave functions of baryons are totally symmetric under the permutation of quarks, 
we have that $(a)+(b)+(c)=3(a)=3(b)=3(c)$~\cite{Schlumpf:1992ce}. 
As an illustration, we only present the calculation for  the diagram (c), 
which contains simpler and cleaner forms with the notation $(q_{\perp},Q_{\perp},\xi,\eta)$.
We can extract the form factors from the matrix elements through the relations
\begin{eqnarray}
	f_1(k^2)&=&\frac{1}{2P^+}\langle {\bf B}_f,P',\uparrow|\bar{q}\gamma^{+}c|{\bf B}_i,P,\uparrow\rangle\,, \nonumber \\
	f_2(k^2)&=&\frac{1}{2P^+}\frac{M_{{\bf B}_i}}{k_{\perp}}\langle {\bf B}_f,P',\uparrow|\bar{q}\gamma^{+}c|{\bf B}_i,P,\downarrow\rangle\,, \nonumber \\
	g_1(k^2)&=&\frac{1}{2P^+}\langle {\bf B}_f,P',\uparrow|\bar{q}\gamma^{+}\gamma_5c|{\bf B}_i,P,\uparrow\rangle\,, \nonumber \\
	g_2(k^2)&=&\frac{1}{2P^+}\frac{M_{{\bf B}_i}}{k_{\perp}}\langle {\bf B}_f,P',\uparrow|\bar{q}\gamma^{+}\gamma_5|{\bf B}_i,P,\downarrow\rangle  \,.\label{fm}
\end{eqnarray}
Note that $f_3$ and $g_3$ cannot be obtained when $k^+=0$, but they are negligible because of the suppressions of the $k^2$ factors.
In addition, the terms associated with $f_3$ and $g_3$ do not contribute to the semileptonic decays~\cite{Geng:2019bfz}.
As a result,  we set both $f_3$ and $g_3$  to be 0 in this study.
With the help of the  momentum distribution functions and the Melosh transformation matrix,  
the
transition matrix elements  can be expressed as
\begin{eqnarray}
&&\langle {\bf B}_f,P',S',S'_z|\bar{q}\gamma^{+}c|{\bf B}_i,P,S,S_z\rangle \nonumber \\
&&=\frac{1}{2^2(2\pi)^6}\int d\xi d\eta d^2q_\perp d^2Q_\perp\Phi(q'_{\perp},\xi,Q'_\perp,\eta)\Phi(q_{\perp},\xi,Q_\perp,\eta)F^{def} F_{abc}\delta_{d}^{a}\delta_{e}^{b}\nonumber\\
&&\times \sum_{s_1,s_2,s_3}\sum_{s'_1,s'_2,s'_3}\langle S',S'_z|s'_1,s'_2,s'_3\rangle\langle s_1,s_2,s_3|S,S_z\rangle
\langle s'_1|R'_1R^{\dagger}_1|s_1\rangle\langle s'_2|R'_2R^{\dagger}_2|s_2\rangle \nonumber\\
&&\times 2P^+\sum_{\lambda'_3\lambda_3}\langle s'_3|R'_3|\lambda'_3\rangle (\delta_{q_{f}q}3\delta_{\lambda'_3\lambda_3}\delta_{cq^c})\langle\lambda_3|R_3^\dagger|s_3\rangle \label{vf}\,,
\end{eqnarray}
\begin{eqnarray}
&&\langle {\bf B}_f,P',S',S'_z|\bar{q}\gamma^{+}\gamma_5c|{\bf B}_i,P,S,S_z\rangle \nonumber \\
&&=\frac{1}{2^2(2\pi)^6}\int d\xi d\eta d^2q_\perp d^2Q_\perp\Phi(q'_{\perp},\xi,Q'_\perp,\eta)\Phi(q_{\perp},\xi,Q_\perp,\eta)F^{def} F_{abc}\delta_{d}^{a}\delta_{e}^{b}\nonumber\\
&&\times \sum_{s_1,s_2,s_3}\sum_{s'_1,s'_2,s'_3}\langle S',S'_z|s'_1,s'_2,s'_3\rangle\langle s_1,s_2,s_3|S,S_z\rangle
\langle s'_1|R'_1R^{\dagger}_1|s_1\rangle\langle s'_2|R'_2R^{\dagger}_2|s_2\rangle \nonumber\\
&&\times 2P^+\sum_{\lambda'_3\lambda_3}\langle s'_3|R'_3|\lambda'_3\rangle (\delta_{q_{f}q}3(\sigma_z)_{\lambda'_3\lambda_3}\delta_{cq^c})\langle\lambda_3|R_3^\dagger|s_3\rangle \label{af}
\end{eqnarray}
Using Eqs.~(\ref{fm}), (\ref{vf}) and (\ref{af}), we find that 
\begin{eqnarray}
\label{f1}
&&f_1(k^2)=\frac{3}{2^2(2\pi)^6}\int d\xi d\eta d^2q_\perp d^2Q_\perp\Phi(q'_{\perp},\xi,Q'_\perp,\eta)\Phi(q_{\perp},\xi,Q_\perp,\eta)(F^{def} F_{abc}\delta_{q_{f}q}\delta_{cq^c}\delta_{d}^{a}\delta_{e}^{b})\nonumber\\
&&\times \sum_{s_1,s_2,s_3}\sum_{s'_1,s'_2,s'_3}\langle S',\uparrow|s'_1,s'_2,s'_3\rangle\langle s_1,s_2,s_3|S,\uparrow\rangle
\prod_{i=1,2,3}\langle s'_i|R'_iR^{\dagger}_i|s_i\rangle \,,
\end{eqnarray}
\begin{eqnarray}
\label{g1}
&&g_1(k^2)=\frac{3}{2^2(2\pi)^6}\int d\xi d\eta d^2q_\perp d^2Q_\perp\Phi(q'_{\perp},\xi,Q'_\perp,\eta)\Phi(q_{\perp},\xi,Q_\perp,\eta)(F^{def} F_{abc}\delta_{q_{f}q}\delta_{cq^c}\delta_{d}^{a}\delta_{e}^{b})\nonumber\\
&&\times \sum_{s_1,s_2,s_3}\sum_{s'_1,s'_2,s'_3}\langle S',\uparrow|s'_1,s'_2,s'_3\rangle\langle s_1,s_2,s_3|S,\uparrow\rangle
\prod_{i=1,2}\langle s'_i|R'_iR^{\dagger}_i|s_i\rangle \langle s'_3|R'_3\sigma_zR^{\dagger}_3|s_3\rangle\,,
\end{eqnarray}
\begin{eqnarray}
\label{f2}
&&f_2(k^2)=\frac{3}{2^2(2\pi)^6}\frac{M_{{\bf B}_i}}{k_{\perp}}\int d\xi d\eta d^2q_\perp d^2Q_\perp\Phi(q'_{\perp},\xi,Q'_\perp,\eta)\Phi(q_{\perp},\xi,Q_\perp,\eta)(F^{def} F_{abc}\delta_{q_{f}q}\delta_{cq^c}\delta_{d}^{a}\delta_{e}^{b})\nonumber\\
&&\times \sum_{s_1,s_2,s_3}\sum_{s'_1,s'_2,s'_3}\langle S',\uparrow|s'_1,s'_2,s'_3\rangle\langle s_1,s_2,s_3|S,\downarrow\rangle
\prod_{i=1,2,3}\langle s'_i|R'_iR^{\dagger}_i|s_i\rangle  \,,
\end{eqnarray}
\begin{eqnarray}
&&g_2(k^2)=\frac{3}{2^2(2\pi)^6}\frac{M_{{\bf B}_i}}{k_{\perp}}\int d\xi d\eta d^2q_\perp d^2Q_\perp\Phi(q'_{\perp},\xi,Q'_\perp,\eta)\Phi(q_{\perp},\xi,Q_\perp,\eta)(F^{def} F_{abc}\delta_{q_{f}q}\delta_{cq^c}\delta_{d}^{a}\delta_{e}^{b})\nonumber\\
&&\times \sum_{s_1,s_2,s_3}\sum_{s'_1,s'_2,s'_3}\langle S',\uparrow|s'_1,s'_2,s'_3\rangle\langle s_1,s_2,s_3|S,\downarrow\rangle
\prod_{i=1,2}\langle s'_i|R'_iR^{\dagger}_i|s_i\rangle \langle s'_3|R'_3\sigma_zR^{\dagger}_3|s_3\rangle\,.
\label{g2}  
\end{eqnarray}

\section{Baryonic transition form factors in MBM}
The formalism  for MBM can be found in Ref.~\cite{PerezMarcial:1989yh}. 
In the calculation of MBM, we take the same notations
as those in Ref.~\cite{PerezMarcial:1989yh}.
In this approach,  the current quark masses are used, given by
\begin{equation}
m_{u,d}= 0.005~\text{GeV}\,,~~~~m_s= 0.28~\text{GeV}\,,~~~~m_c = 1.5~ \text{GeV}\,,~~~~R=5~\text{GeV}^{-1}\,,
\end{equation}
where $R$ corresponds to the bag size, which is valid at least for the charmed baryons~\cite{bag1,bag2,bag3,bag4}.
Note that the form factors can only be evaluated at $\vec{k}=0$ ($k^2=(M_1-M_2)^2$). 
For $\vec{k}\neq0$, the bag is not at the rest frame of the initial baryon, and we will face the problem of how to boost the state in MBM, which is very subtle and beyond the study of this work~\cite{Boosting the bag}.
The form factors are decomposed as follows:
\begin{eqnarray}
f_{1}&=&\mathcal{V}_{0}-\mathcal{V}_{M} \Delta M^{2} / M_{12}-\mathcal{V}_{V} \Delta M \,,\nonumber\\
f_{2}&=&\left(-\mathcal{V}_{0}+\mathcal{V}_{M} M_{12}+\mathcal{V}_V \Delta M\right) M_{1} / M_{12}\,, \nonumber\\
g_{1}&=&\left(1-\Delta M^{2} / 2 M_{12}^{2}\right) \mathcal{A}_{s}+\left(\mathcal{A}_{T} \Delta M-\mathcal{A}_{0}\right) 4 M_{1} M_{2} \Delta M / M_{12}^{2}\,,\nonumber \\
g_{2}&=&\left(\mathcal{A}_{T} \Delta M-\mathcal{A}_{s} \Delta M / 8 M_{1} M_{2}-\mathcal{A}_{0}\right) 4 M_{1}^{2} M_{2} / M_{12}^{2} 
\end{eqnarray}
with $\Delta M = M_1- M_2$, $M_{12}=M_1 + M_2$ and 
\begin{eqnarray}
\mathcal{V}_{0}&=&A R^{3}\left(W^{i}_+W^{f}_+I_{00}+W^{i}_-W^{f}_-I_{11}\right)\,, \nonumber\\
\mathcal{V}_{}&=&A  R^{3}\left(W^{i}_-W^{f}_+I_{10}-W^{i}_+W^{f}_-I_{01}\right)(R / 3)\,, \nonumber\\
\mathcal{V}_{M}&=&A R^{3}\left(W^{i}_-W^{f}_+I_{10}+W^{i}_+W^{f}_-I_{01}\right)(R / 3)\,, \nonumber\\
\mathcal{A}_{0}&=&A R^{3}\left(W^{i}_-W^{f}_+I_{10}-W^{i}_+W^{f}_-I_{01}\right)(R / 3)\,, \nonumber\\
\mathcal{A}_s&=&AR^3\left(W^i_+W^f_+ - W^i_-W^f_-I_{11}/3\right) \,,\nonumber\\
\mathcal{A}_{T}&=&A  R^{3} W^{i}_-W^{f}_- J_{11}(-2 R^2 / 15)\,,
\end{eqnarray}
where $A$ is the normalized factor for the baryon, corresponding to the baryon spin-flavor structures 
in Table~II of Ref.~\cite{PerezMarcial:1989yh}, 
$W^q_{\pm}$ are associated with the normalized factors for quarks, given by
\begin{equation}
W_{\pm}^{q} \equiv\left(\frac{\omega^{q}\pm m_{q}}{\omega^{q}}\right)^{1 / 2}
\end{equation} 
with  $q=i$ or $f$  the quark flavor and $\omega^q$  the quark energy, 
and $I$ and $J$ stand for
the overlap factors for the quark wave functions, defined by
\begin{eqnarray}
I_{n n} &\equiv& \int_{0}^{1} d t t^{2} j_{n}\left(t x_{0}^{i}\right) j_{n}\left(t x_{0}^{f}\right), \quad n=0,1\nonumber\\
I_{n m} &\equiv &\int_{0}^{1} d t t^{3} j_{n}\left(t x_{0}^{i}\right) j_{m}\left(t x_{0}^{f}\right), \quad n, m=0,1~(n \neq m)\nonumber\\
J_{11} &\equiv& \int_{0}^{1} d t\, t^{4} j_{1}\left(t x_{0}^{i}\right) j_{1}\left(t x_{0}^{f}\right)\,,
\end{eqnarray}
with $j_n$  the Bessel function and $x^q_0$  the lowest root of the transcendental equation of
\begin{eqnarray}
\tan(x^q)=\frac{x^q}{1-m_qR-[(x^q)^2+(m_qR)^2]^{1/2}}\,.
\end{eqnarray}

\section{Numerical Results}
As shown in Sec.~II, the baryonic transition form factors in LFCQM can be evaluated only in the space-like region ($k^2=-k^2_\perp$) because of the condition $k^+=0$. Thus, we follow the standard procedures in Refs.~\cite{Cheng:2003sm,Ke:2007tg,Cheng:2004cc} to extract the information of the form factors in the time-like region.
These procedures have widely been tested and discussed in the mesonic sector~\cite{Jaus:1991cy,Jaus:1996np}.
 We fit $f_{1(2)}(k^2)$ and $g_{1(2)}(k^2)$ with some analytic functions in the space-like region, which are
 analytically continued to the physical time-like region $(k^2>0)$.  We employ the numerical values of the constituent quark masses and shape parameters in Table~I. 
 \begin{table}
\label{sh}
\caption{Values of the constituent quark masses ($m_i$) and shape parameters ($\beta_{q{\bf B}}$ and $\beta_{{Q\bf B}}$) in units of GeV,
where $\beta_{I,II}=\beta_{Qn}=\beta_{qn}$.} 
\begin{tabular}{ccccccccc}
\hline
$m_c$&$m_s$&$m_d$&$m_c$&$\beta_{q\Lambda_c}$&$\beta_{Q\Lambda_c}$&$\beta_{q\Lambda}$&$\beta_{Q\Lambda}$& $\beta_{I,II}$\\
\hline
1.3&0.4&0.26&0.26&0.44&0.54&0.44&0.37&0.22,0.44\\
\hline
\end{tabular}
\end{table} 
 The values of the shape parameters can be determined approximately by the calculations in the mesonic sectors~\cite{Chang:2018zjq,Ke:2019smy}.
 Since the strength of the quark-quark ($qq'$) potential is a half of the quark-antiquark  $q\bar{q}'$ one, the shape parameters of the quark pairs $\beta_{q\bf B}$ should be $\sqrt{2}$ smaller than those in the mesonic sectors~\cite{Ke:2019smy}.
  Meanwhile, the reciprocals of the shape parameters are related to the sizes of systems. Consequently,  
  we adopt $\beta_{q\Lambda_{(c)}}\simeq 2(\beta_{u\bar{d}}/\sqrt{2})$,
   where the factor of 2 is used to parameterize the effects of the diquark clusterings, resulting in the light quark pairs to be more compact.  
   For the quark-diquark shape parameters $\beta_{Q\Lambda_{(c)}}$, we choose the values 
   of $\beta_{{s(c)\bar{s}}}$  without any additional factors.  The diquark cluster effectively forms a color anti-triplet, and hence shares the same potential strength as the $q\bar{q}'$ one. Finally, because of the isospin symmetry, all constitute quarks in the neutron are expected to have
    the same momentum distribution, so that the shape parameters are $\beta_{Qn}=\beta_{qn}=\beta$.  
    We use two scenarios for $\beta$ to describe the  quarks in the neutron.  The first one is from the harmonic oscillator picture, which leads to
    $\beta_I=\beta_{u\bar{d}}/\sqrt{2}\simeq0.22\text{ GeV}$ through the quark-quark interaction, in which the value of $0.22\text{ GeV}$ is consistent with $R=5\text{ GeV}^{-1}$ in the MIT bag model. The other one is to maintain the shape parameters of  $\beta_{q\Lambda_{(c)}}$ to be the same, 
    i.e. $\beta_{II}=\beta_{Qn}=\beta_{qn}=\beta_{q\Lambda_{c}}=0.44 \text{ GeV}$ .
  By using  Eqs.~(\ref{f1})-(\ref{g2}), we compute totally 32 points for all form factors from $k^2=0$ to $k^2=-9.7\text{ GeV}^2$.
With the MATLAB curve fitting toolbox,
  we present our results of  $\Lambda_c^+\to \Lambda$ in Figs.~\ref{lclvffit} and \ref{lclaffit} and $\Lambda_c^+\to n$ in Figs.~\ref{lcnvffit} and \ref{lcnaffit}
   based on  $95\%$ confidence bounds  given in Appendix, respectively.
To fit the $k^2$ dependences of the form factors in the space-like region, we  use the form   
\begin{eqnarray}
	F(k^2)=\frac{F(0)}{1-q_1k^2+q_2k^4} \,.
\end{eqnarray}
We present our fitting results in Table~\ref{nf}.
\begin{table}
	\caption{Fitting results of the form factors in LFCQM, where (I) and (II) represent the two scenarios of $\beta_I=0.22$ and $\beta_{II}=0.44$
	for $\Lambda_c^+ \to n$, respectively.
	}
\begin{tabular}{ccccc}
	\hline
	\hline
	\multicolumn{5}{c}{$\Lambda_c^+ \to \Lambda$}\\
	\hline
	&$f_1$&$f_2$&$g_1$&$g_2$\\
	$F(0)$&$0.67\pm0.01$&$0.76\pm0.02$&$0.59\pm0.01$&$-(1.59\pm0.05)\times10^{-3}$\\
	$q_1$ (GeV$^{-2}$)&$1.45\pm0.29$&$1.42\pm0.29$&$1.198\pm0.26$&$0.53\pm0.24$\\
	$q_2$ (GeV$^{-4}$)&$2.39\pm0.45$&$2.34\pm0.44$&$1.904\pm0.36$&$1.03\pm0.23$\\
		\hline
	\multicolumn{5}{c}{$\Lambda_c^+ \to n$}\\
	\hline
	(I) &$f_1$&$f_2$&$g_1$&$g_2$\\
	$F(0)$&$0.34\pm0.01$&$0.40\pm0.01$&$0.30\pm0.01$&$-0.14\pm0.01$\\
	$q_1$ (GeV$^{-2}$)&$1.79\pm0.36$&$1.83\pm0.37$&$1.56\pm0.33$&$2.08\pm0.41$\\
	$q_2$ (GeV$^{-4}$) &$3.59\pm0.68$&$3.65\pm0.69$&$3.03\pm0.56$&$4.24\pm0.83$\\
	\hline
	(II) &$f_1$&$f_2$&$g_1$&$g_2$\\
	$F(0)$&$0.83\pm0.02$&$1.05\pm0.02$&$0.71\pm0.02$&$0.27\pm0.01$\\
	$q_1$ (GeV$^{-2}$)&$1.25\pm0.26$&$1.20\pm0.25$&$0.94\pm0.22$&$1.37\pm0.27$\\
	$q_2$ (GeV$^{-4}$)&$1.85\pm0.34$&$1.77\pm0.33$&$1.36\pm0.25$&$2.08\pm0.28$\\
	\hline
	\hline
\end{tabular}
\label{nf}
\end{table}

For MBM, we assume the $k^2$ dependence of the form factors as follows:
\begin{eqnarray}
\label{FG}
f_{i}(k^2)=\frac{(1+d_f)f_{i}(0)}{(1-\frac{k^2}{M^2_V})^2+d_f}
\nonumber \\
g_{i}(k^2)=\frac{(1+d_g)g_{i}(0)}{(1-\frac{k^2}{M^2_A})^2+d_g}
\end{eqnarray}
where $M_V=2.112~(2.010)$~GeV and $M_A=2.556~(2.423)$~GeV, 
while $d_f$ and $d_g$ are fitted to be $0.2$ and $d_g=0.1$, respectively. 
We will call the $k^2$-dependent forms in Eq.~(\ref{FG}) as the Lorentzian type.
We list $f_{i}(0)=f_i$ and $g_{i}(0)=g_i$ in Table~\ref{MBM}.
\begin{table}
	\caption{Fitting results of the form factors in MBM}
	\begin{tabular}{ccccc}
	\hline\hline
		&$f_1$&$f_2$&$g_1$&$g_2$\\
		\hline
		$\Lambda_c^+\to \Lambda$&0.54&0.22&0.52&-0.06\\
		$\Lambda_c^+\to n$&0.40&0.22&0.43&-0.07\\
		\hline\hline
	\end{tabular}
\label{MBM}
\end{table}

In order to calculate the decay branching ratios and other physical quantities, we introduce the
the helicity amplitudes of  $H^{V(A)}_{\lambda_2\lambda_W}$,  which give more intuitive physical pictures and simpler expressions when discussing 
the asymmetries of the decay processes, such as   the integrated (averaged) asymmetry, also known as the longitudinal polarization of the daughter baryon.
Relations between the helicity amplitudes and form factors are given by
\begin{eqnarray}
H^{V}_{\frac{1}{2}1}&=&\sqrt{2K_-}\left(-f_1-\frac{M_{\bf B_{i}}+M_{\bf B_{f}}}{M_{\bf B_{i}}}f_2 \right) \,,\nonumber \\
H^{V}_{\frac{1}{2}0}&=&\frac{\sqrt{K_-}}{\sqrt{k^2}}\left((M_{\bf B_{i}}+M_{\bf B_{f}})f_1+\frac{k^2}{M_{\bf B_{i}}}f_2\right)\,,\nonumber \\
H^{V}_{\frac{1}{2}t}&=&\frac{\sqrt{K_+}}{\sqrt{k^2}}\left((M_{\bf B_{i}}+M_{\bf B_{f}})f_1+\frac{k^2}{M_{\bf B_{i}}}f_3\right)\,,\nonumber \\
H^{A}_{\frac{1}{2}1}&=&\sqrt{2K_+}\left(g_1-\frac{M_{\bf B_{i}}-M_{\bf B_{f}}}{M_{\bf B_{i}}}g_2\right)\,,\nonumber \\
H^{A}_{\frac{1}{2}0}&=&\frac{\sqrt{K_+}}{\sqrt{k^2}}\left(-(M_{\bf B_{i}}-M_{\bf B_{f}})g_1+\frac{k^2}{M_{\bf B_{i}}}g_2\right) \,,\nonumber\\
H^{A}_{\frac{1}{2}t}&=&\frac{\sqrt{K_-}}{\sqrt{k^2}}\left(-(M_{\bf B_{i}}-M_{\bf B_{f}})g_1+\frac{k^2}{M_{\bf B_{i}}}g_3\right)\,,
\label{amp}
\end{eqnarray} 
where $K_{\pm}=(M_{\bf B_i}-M_{\bf B_f})^2-k^2$. 

The differential decay widths and asymmetries can be expressed in the analytic forms in terms of the helicity amplitudes, which can be found 
in  Ref.~\cite{Geng:2019bfz}.
In our numerical calculations,
we  use the center value of  $\tau_{\Lambda_c^+}=203.5\times 10^{-15}s$ in Eq.~(\ref{q0})~\cite{Aaij:2019lwg}.
Our
predictions of the decay branching ratios (${\cal B}$s) and asymmetries ($\alpha$s) are listed in Table~\ref{result}.
In Table~\ref{com}, we compare our results with the experimental data 
 and  various other calculations in the literature.

\begin{table}[h]
	\caption{Predictions of the decay branching ratios and asymmetry parameters in LFCQM and  MBM, 
where (I) and (II) represent the two scenarios of $\beta=0.22$ and 0.44
	for $\Lambda_c^+ \to n$, respectively.	
	}
	\begin{tabular}{c|cc|cc}
	\hline\hline
		&\multicolumn{2}{c|}{LFCQM}&\multicolumn{2}{c}{MBM}\\
		\hline
		&${\cal B}(\%)$&$\alpha$&${\cal B}(\%)$&$\alpha$\\
		$\Lambda_c^+\to \Lambda e^+ \nu_e$&$3.36\pm0.87$&$-0.97\pm0.03$&$3.48$&$-0.83$\\
		$\Lambda_c^+\to \Lambda \mu^+ \nu_\mu$&$3.21\pm0.85$&$-0.97\pm0.03$&$3.38$&$-0.82$\\
	\multirow{2}{*}{$\Lambda_c^+\to ne^+\nu_e$}&$0.057\pm0.015$ (I) &$-0.98\pm0.02$ (I)&\multirow{2}{*}{$0.279$}&\multirow{2}{*}{$-0.87$}\\
	&$0.36\pm0.15$ (II)&$-0.96\pm0.04$ (II)&&\\
		\multirow{2}{*}{$\Lambda_c^+\to n\mu^+\nu_\mu$}&$0.054\pm0.015$ (I)&$-0.98\pm0.01$ (I)&\multirow{2}{*}{$0.273$}&\multirow{2}{*}{$-0.87$}\\
		&$0.34\pm0.14$ (II)&$-0.96\pm0.04$ (II)&&\\
		\hline\hline
	\end{tabular}
	\label{result}
\end{table}

\begin{table}[h]
	\caption{ Our results in comparisons with the experimental data and those in various calculations in the literature.}
	\label{com}
	{	
		\begin{tabular}{c|cc|cc}
		\hline\hline
			&\multicolumn{2}{c|}{$\Lambda_c^+\to \Lambda e^+\nu_e$}&\multicolumn{2}{c}{$\Lambda_c^+\to n e^+\nu_e$}\\
			\hline
			&${\cal B}(\%)$&$\alpha$&${\cal B}(\%)$&$\alpha$\\
		\multirow{2}{*}{LFCQM}&\multirow{2}{*}{$3.36\pm0.87$}&\multirow{2}{*}{$-0.97\pm0.03$}&$0.057\pm0.015$(I)&$-0.98\pm0.01$ (I)\\
			&&&$0.36\pm0.15$ (II)&$-0.96\pm0.04$ (II)\\
			MBM&$3.48$&$-0.83$&$0.279$&$-0.87$\\
			Data~\cite{Tanabashi:2018oca}&$3.6\pm0.4$& $-0.86\pm0.04$&-&-\\
			$SU(3)$~\cite{Geng:2019bfz}&$3.2\pm0.3$&$-0.86\pm0.04$&$0.51\pm0.04$&$-0.89\pm0.04$\\
			HQET~\cite{Cheng:1995fe}&1.42&-&-&-\\
			LF~\cite{Zhao:2018zcb}&1.63&-&0.201&-\\
			MBM~\footnote[3]{Although the values of $f_i$ and $g_i$ are the same at the zero recoil point ($\vec{q}=0$), we use the Lorentzian type of the
			$k^2$ dependences for the form factors instead of the  dipole ones in this work.}~\cite{PerezMarcial:1989yh}
			&2.6&-&0.20&-\\
			NRQM~\cite{PerezMarcial:1989yh}&3.2&-&0.30&-\\
			LQCD~\cite{Meinel:2016dqj,Meinel:2017ggx}&$3.80\pm0.22$&-&$0.410\pm0.029$&-\\
			RQM~\cite{Faustov:2016yza}&3.25&-&0.268&-\\
			CCQM~\cite{Gutsche:2014zna,Gutsche:2015rrt}&2.78&-0.87&0.202&-\\
		\hline\hline	
		\end{tabular}
	}
\end{table}

In LF~\cite{Zhao:2018zcb} and the heavy effective theory (HQET)~\cite{Cheng:1995fe}, the authors use a specific spin-flavor structure of $c(ud-du)\chi_{s_z}^{\rho_3}$ 
for the charmed baryon state,
in which only the permutation relation is considered  between light quarks. In addition, they assume that 
the diquarks from the light quark pairs are spectators and structureless. These  simplifications 
make the  results 
in Refs.~\cite{Zhao:2018zcb,Cheng:1995fe}
to be not good compared with the  experimental data as shown in Table~\ref{com}.
 Based on the Fermi statistics, the overall spin-flavor-momentum structures are determined, from which 
  the parameters like quark masses, baryon masses and shape parameters can recover the spin-flavor symmetry. 
It is  interesting  to see  that when we consider the scenario II in the neutron, the same shape parameters of $\beta_{{Q\bf B}}$ and $\beta_{q{\bf B}}$
 in our study  imply the totally symmetric momentum distribution of three constituent quarks in the baryon. 
In addition,
the flavor symmetry breaking effect due to the quark masses seems to get canceled due to the clustering effect of the shape parameters in 
the momentum distribution functions.  
Our numerical results indicate  that the form factors follow the Lorentzian functions of
$F(k^2)=F(0)/(1-q_1k^2+q_2k^4)$ in both $\Lambda_c^+ \to \Lambda(n)$ processes. 
Our results of  $f_i(k^2)\neq g_i(k^2)$ show that
 the heavy quark symmetry is broken  because the constituent charm quark mass is not heavy enough.
 
 From Table~\ref{result}, we predict  that   ${\cal B}(\Lambda_c^+\to \Lambda e^+\nu_e)=(3.36\pm0.87)\times 10^{-2}$ and $\alpha(\Lambda_c^+\to \Lambda e^+\nu_e)=-0.96\pm0.03$, and ${\cal B}(\Lambda_c^+\to n e^+\nu_e)=(0.57\pm0.15,3.6\pm1.5)\times 10^{-3}$,
 and $\alpha(\Lambda_c^+\to n e^+\nu_e)=(-0.98\pm0.02,-0.96\pm0.04)$ with the two  scenarios of (I) and (II) 
   in LFCQM, in which the  value of  ${\cal B}(\alpha)$ for the mode of $ \Lambda_c^+\to \Lambda e^+\nu_e$ 
  is lower (higher) than but acceptable by the experimental one  $(3.6\pm0.4)\times 10^{-2}~ (-0.86\pm0.04)$ in PDG~\cite{Tanabashi:2018oca}. 
  The errors in our results mainly come from the numerical fits 
  of the MATLAB curve fitting toolbox  given in Appendix, in which the $95\%$ confidence bounds are broadened and tightened in the time-like  
   space-like regions, respectively.
  Our results are also consistent with those  in LQCD~\cite{Meinel:2016dqj,Meinel:2017ggx},
  the relativistic quark model (RQM)~\cite{Faustov:2016yza} and covariant confinement quark model (CCQM)~\cite{Gutsche:2014zna,Gutsche:2015rrt}.  
For MBM, although the semi-leptonic processes have been fully studied in Ref.~\cite{PerezMarcial:1989yh}, their results are mismatched with the current data. 
By using the same formalism with the same input parameters, we are able to  get the same values of the form factors at the zero recoil point.
By taking the Lorentzian $k^2$ dependences for the form factors, inspired from our LF calculations, we obtain much better results
as shown in Table~\ref{com}. 
It is interesting to see that our results for $\Lambda_c^+\to n e^+\nu_e$ in LFCQM with the scenario II is consistent with most of models. One the other hand, the prediction of the scenario I is
much smaller than those in the other calculations. This suppression comes from the wave function mismatching between the diquark in the charmed baryon and ordinary quark pairs in the neutron. 

\section{Conclusions}
We have studied the semi-leptonic decays of $\Lambda_c^+ \to \Lambda(n)\ell^+ \nu_{\ell}$ in the two dynamical approaches of LFCQM and MBM.
We have used  the Fermi statistics to determine the overall spin-flavor-momentum structures 
and recover the spin-flavor symmetry with the quark and baryon masses and shape parameters.
We have found that 
 ${\cal B}( \Lambda_c^+ \to \Lambda e^+ \nu_{e})=(3.36\pm0.87)\%$ and $3.48\%$ in LFCQM and MBM, respectively,
 which are consistent with the experimental data of  $(3.6\pm0.4)\times 10^{-2}$~\cite{Tanabashi:2018oca} as well as
 the values predicted by $SU(3)_F$~\cite{Geng:2019bfz}, LQCD~\cite{Meinel:2016dqj,Meinel:2017ggx},  
 RQM~\cite{Faustov:2016yza} and CCQM~\cite{Gutsche:2014zna,Gutsche:2015rrt}, but about a factor of two larger than those in HQET~\cite{Cheng:1995fe}
  and LF~\cite{Zhao:2018zcb}.
We have also obtained that 
${\alpha}( \Lambda_c^+ \to \Lambda e^+ \nu_{e})=(-0.97\pm0.03)$ and $-0.83$ in LFCQM and MBM, 
which are  lower and higher than but acceptable by the experimental data of  $-0.86\pm0.04$~\cite{Tanabashi:2018oca}, respectively.
 We have  predicted that 
${\cal B}(  \Lambda_c^+ \to n e^+ \nu_{e})=(0.57\pm0.15, 3.6\pm1.5)\times 10^{-3}$ and
${\alpha}( \Lambda_c^+ \to n e^+ \nu_{e})=(-0.98\pm0.02,-0.96\pm0.04)$  with the two different scenarios of (I, II) in LFCQM, and ${\cal B}(  \Lambda_c^+ \to n e^+ \nu_{e})=0.279\times 10^{-2}$ and ${\alpha}( \Lambda_c^+ \to n e^+ \nu_{e})=-0.87$  in MBM,
in which our results of  ${\cal B}(  \Lambda_c^+ \to n e^+ \nu_{e})$ in MBM and LFCQM (II) are consistent with  those
in  RQM~\cite{Faustov:2016yza} and CCQM~\cite{Gutsche:2014zna}, but about two times smaller than the values in $SU(3)_F$~\cite{Geng:2019bfz} and LQCD~\cite{Meinel:2016dqj,Meinel:2017ggx}.
On the other hand, our results of ${\cal B}(  \Lambda_c^+ \to n e^+ \nu_{e})$ in LFCQM (I) is much smaller than other calculations. This additional  suppression could be understood by the wave function mismatching between  the diquark and ordinary quark pairs.
 It is clear that our predicted values for the decay branching ratio and asymmetry in $ \Lambda_c^+ \to n e^+ \nu_{e}$ 
could be tested in the ongoing experiments at BESIII, LHCb and  BELLEII. 
Finally, we remark that
 our  calculations in LFCQM and MBM can be also extended to the other charmed  baryons, such as $\Xi_c^{+,0}$, and even b baryons.
 
\section*{Appendix}

\begin{figure}[h]
	\subfigure{\includegraphics[width=4.5in]{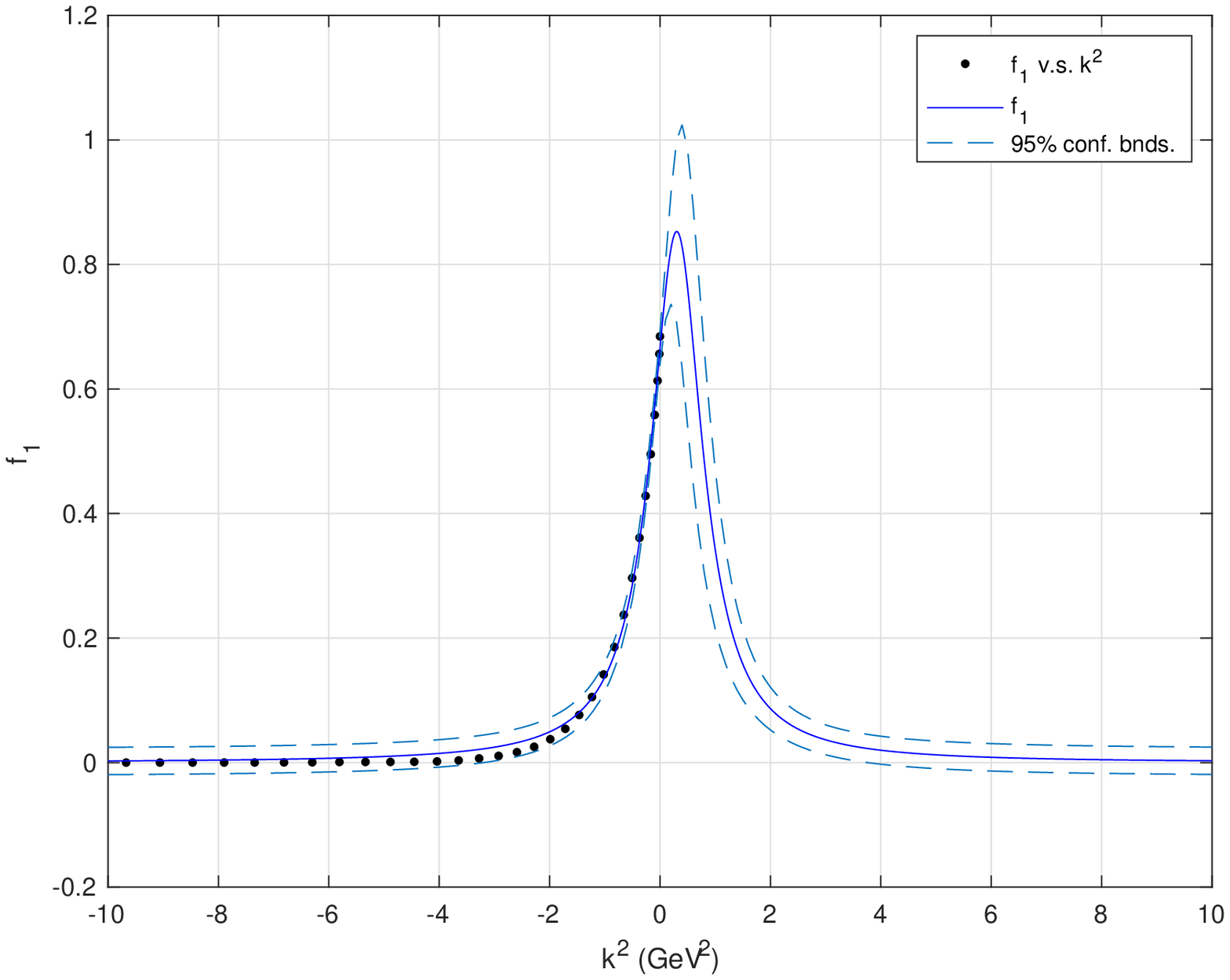}}
	\subfigure{\includegraphics[width=4.5in]{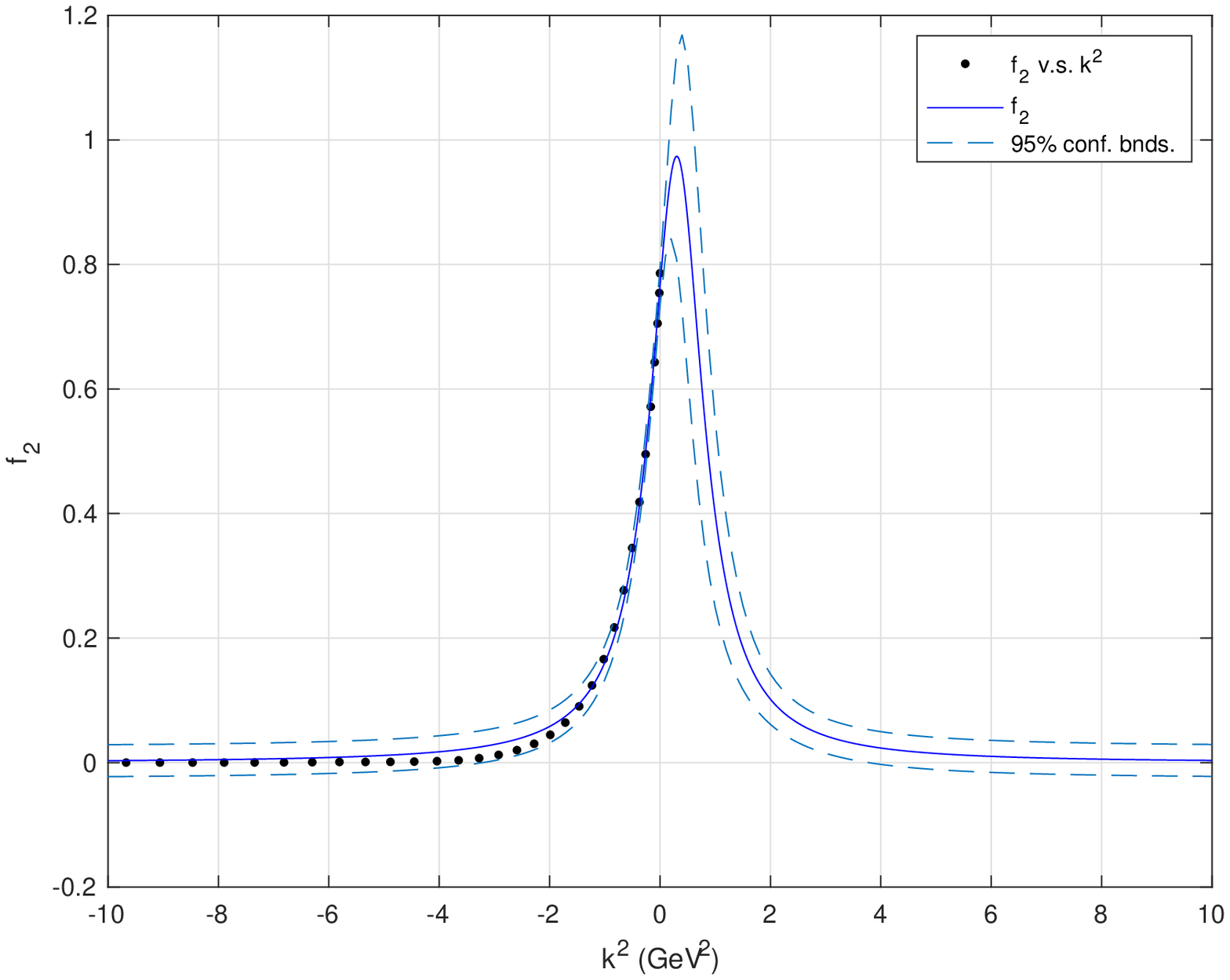}}
	\caption{Vector form factors of $f_{1,2}$ with respect to the transfer momentum $k^2$ in  unit of $\text{ GeV}^2$ for $\Lambda_c^+\to \Lambda$.}
	\label{lclvffit}
\end{figure}

\begin{figure}[h]
	\subfigure{\includegraphics[width=4.5in]{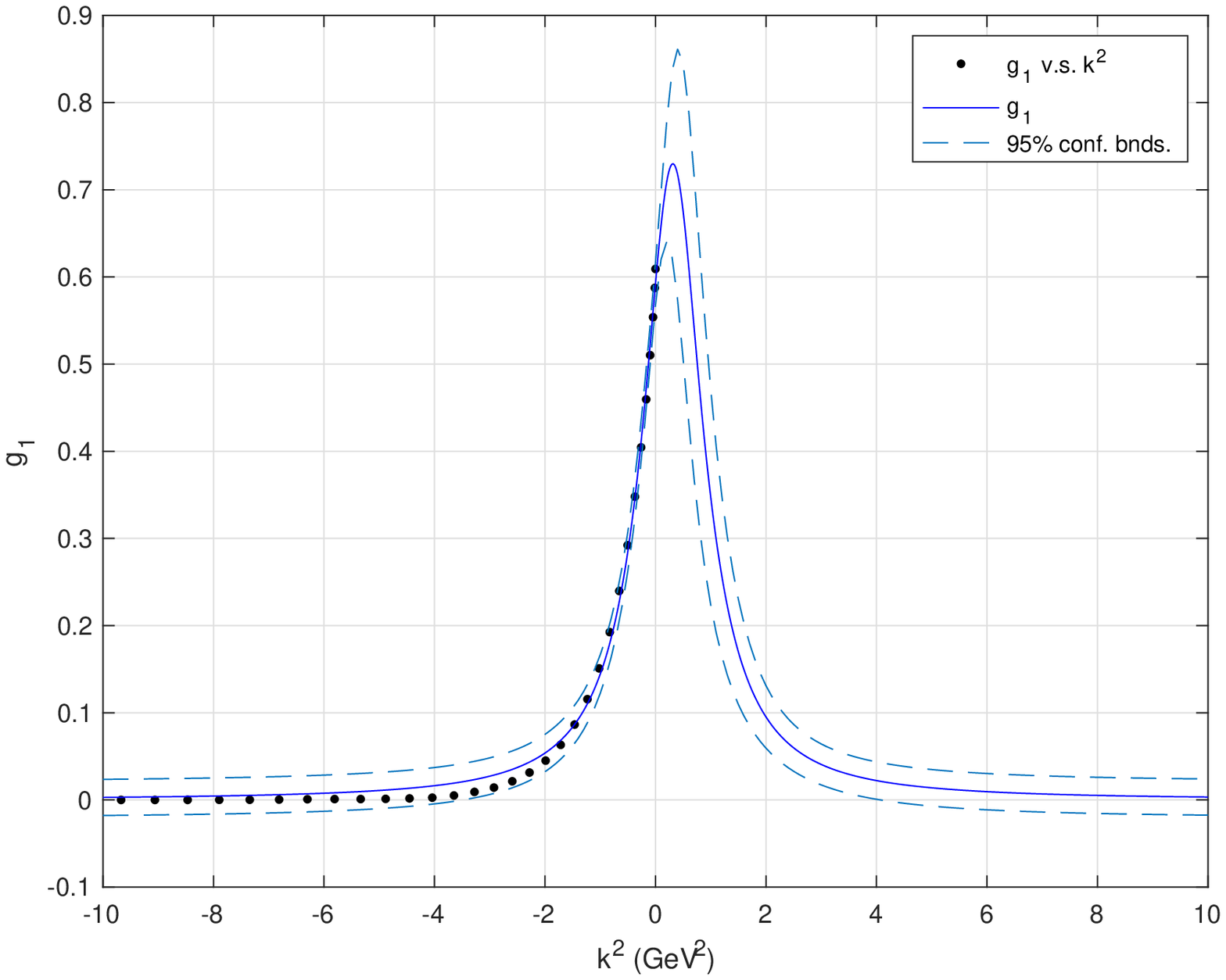}}
	\subfigure{\includegraphics[width=4.5in]{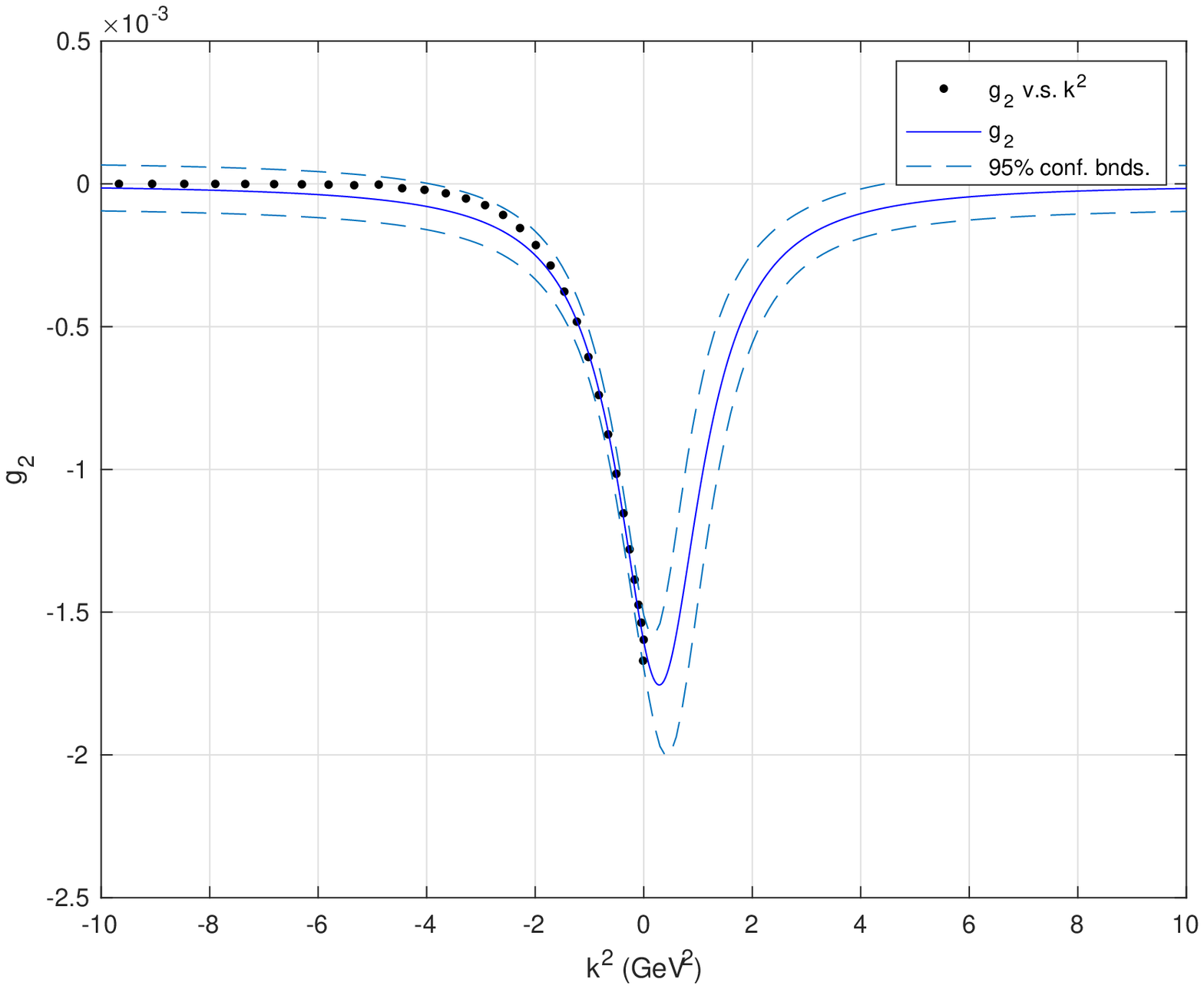}}
	\caption{Axial-vector form factors of $g_{1,2}$ with respect to the transfer momentum $k^2$ in  unit of $\text{ GeV}^2$ in $\Lambda_c^+\to \Lambda$ .}
	\label{lclaffit}
\end{figure}

\begin{figure}[h]
	\subfigure{\includegraphics[width=4.5in]{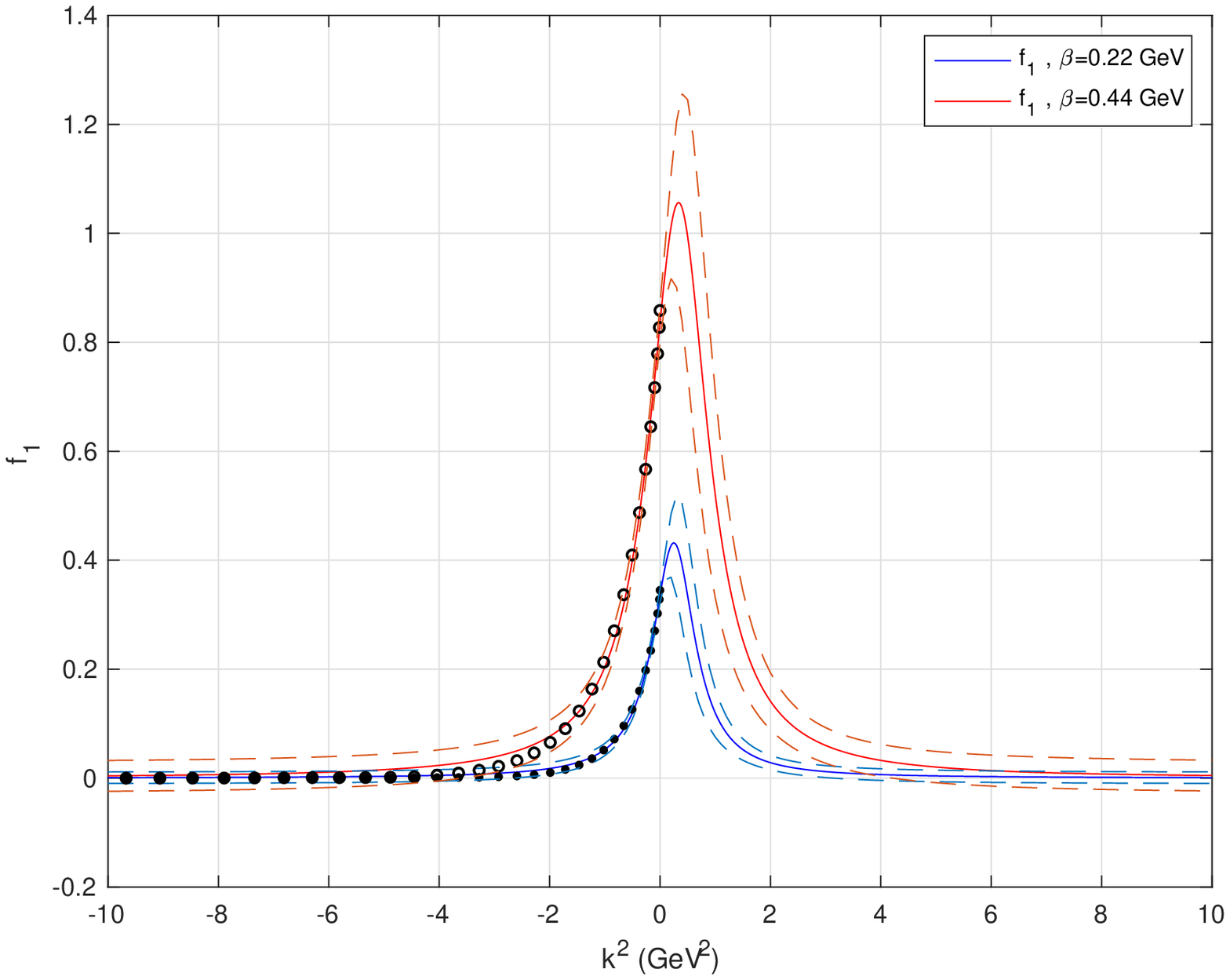}}
	\subfigure{\includegraphics[width=4.5in]{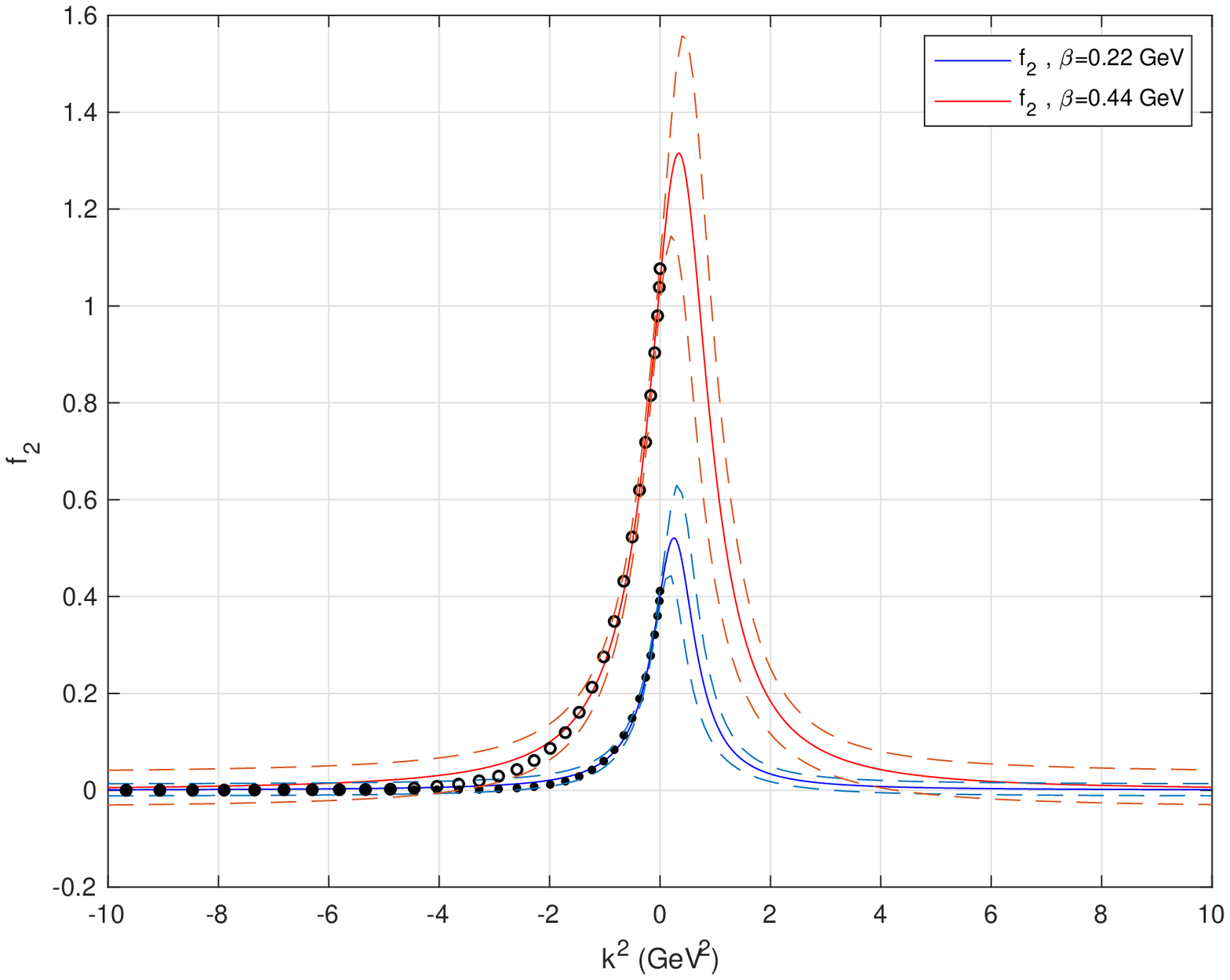}}
	\caption{Legend is the same as Fig.~\ref{lclvffit} but for $\Lambda_c^+\to n$.}
	\label{lcnvffit}
\end{figure}

\begin{figure}[h]
	\subfigure{\includegraphics[width=4.5in]{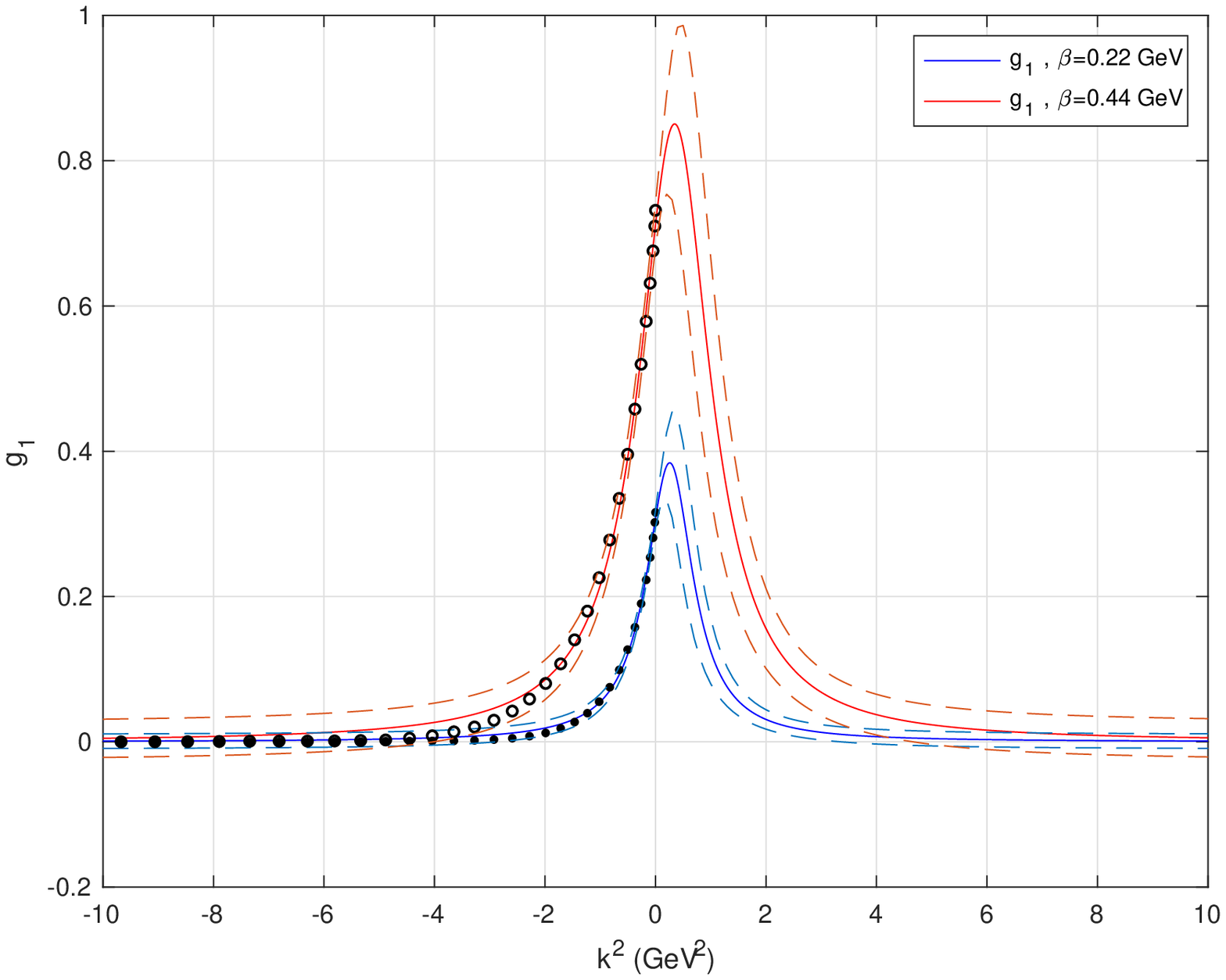}}
	\subfigure{\includegraphics[width=4.5in]{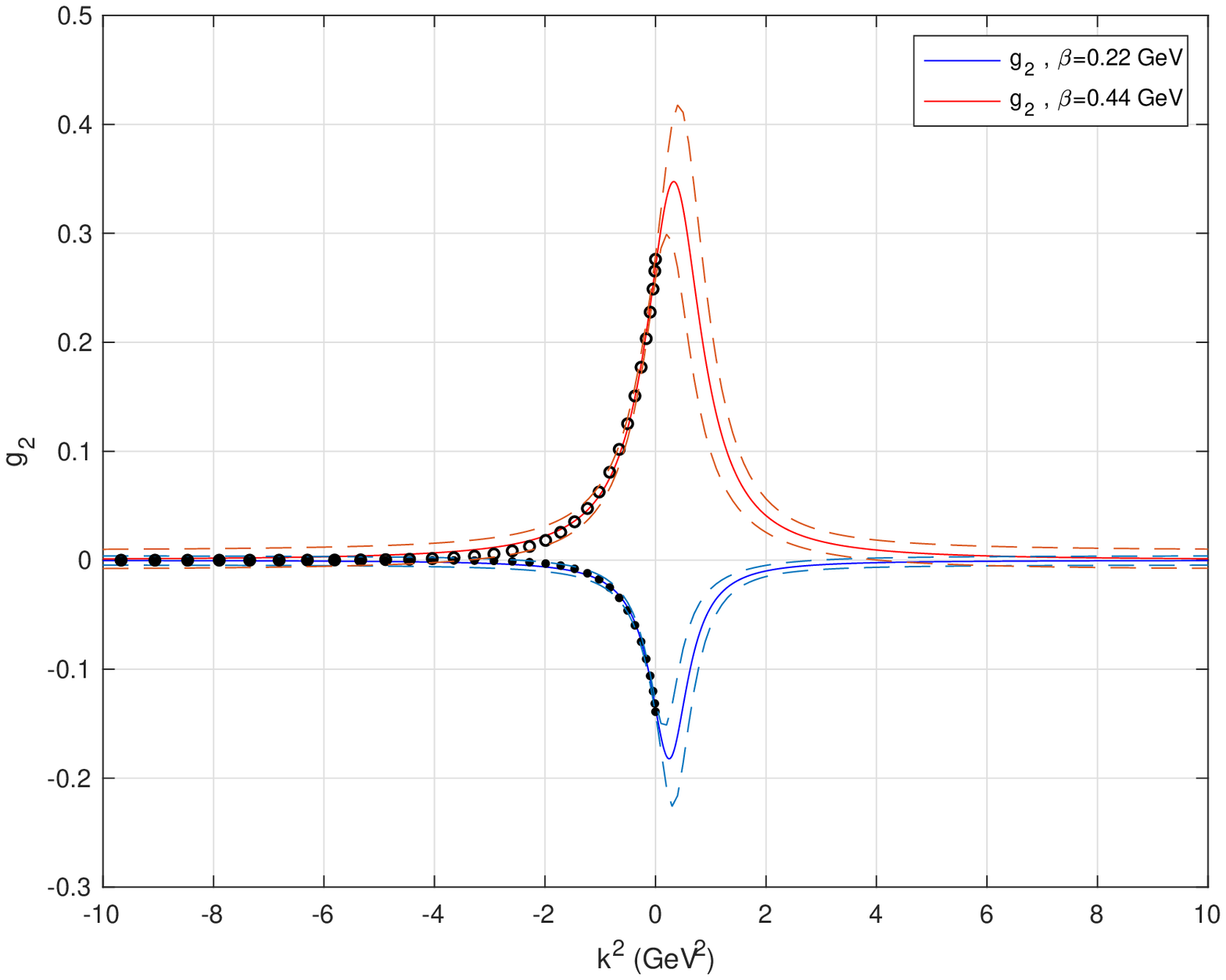}}
	\caption{Legend is the same as Fig.~\ref{lclaffit} but for $\Lambda_c^+\to n$.}
	\label{lcnaffit}
\end{figure}

We now show our numerical results for the form factors in Eqs.~(\ref{f1})-(\ref{g2}) in LFCQM.
In Fig.~\ref{lclvffit},  we plot the vector form factors of $f_{1,2}$ with respect to the transfer momentum $k^2$ in  unit of $\text{ GeV}^2$ for $\Lambda_c^+\to \Lambda$, where the symbol of ``$\bullet$'' denotes the value calculated by Eqs.~(\ref{f1}) and (\ref{f2}) from $k^2=0$ to $-9.7\text{ GeV}^2$ with Mathematica,
while the blue line corresponds to the fitted function by the MATLAB curve fitting toolbox and the dashed line represents
 the $95\%$ confidence bound of the fit.
Similarly, we depict the axial-vector form factors of $g_{1,2}$ in Fig.~\ref{lclaffit}. 
The corresponding results for $\Lambda_c^+\to n$ are given in Figs.~\ref{lcnvffit} and \ref{lcnaffit}.

\section*{ACKNOWLEDGMENTS}
This work was supported in part by National Center for Theoretical Sciences and 
MoST (MoST-107-2119-M-007-013-MY3).

\end{document}